# Recovery of the Fermi-liquid state in $U_3Ni_3Sn_4$ by pressure

P. Estrela, A. de Visser*, and F.R. de Boer

*Van der Waals-Zeeman Institute, University of Amsterdam,*

*Valckenierstraat 65, 1018 XE Amsterdam, The Netherlands*

T. Naka

*National Research Institute for Metals, 1-2-1 Sengen, Tsukuba, Ibaraki 305-0047, Japan*

and

L. Shlyk

*B. Verkin Institute for Low Temperature Physics & Engineering,*

*National Academy of Sciences of Ukraine,*

*47 Lenin Avenue, 310164 Kharkov, Ukraine*

Measurements of the resistivity of single-crystalline $U_3Ni_3Sn_4$ under hydrostatic pressure up to 1.8 GPa are reported. We show that the temperature $T_{FL}$, below which the resistivity obeys the Fermi-liquid expression $\rho \sim T^2$, increases with pressure as $T_{FL} \sim (p-p_c)^{1/2}$. The analysis of the data within the magnetotransport theory of Rosch lends strong support for an antiferromagnetic quantum critical point in $U_3Ni_3Sn_4$ at a negative critical pressure $p_c$ of -0.04±0.02 GPa. The proximity to an antiferromagnetic instability in $U_3Ni_3Sn_4$ is consistent with the previously reported non-Fermi liquid term in the low-temperature specific heat.

Corresponding author:  Dr. A. de Visser
                                 Van der Waals-Zeeman Institute
                                 University of Amsterdam
                                 Valckenierstraat 65, 1018 XE Amsterdam
                                 The Netherlands
                                 Phone: +31-20-5255732
                                 Fax: +31-20-5255788
                                 E-mail: devisser@science.uva.nl

The family of ternary isostructural cubic stannides $U_3T_3Sn_4$, where T = Ni, Cu, Pt or Au, present a promising series of compounds to study $5f$ electron-correlation effects [1]. Moderate heavy-electron behavior is found for all four compounds at low temperatures, which affirms the dominant role of $5f$ electron hybridization phenomena in the physical properties. Specific-heat experiments, performed down to a temperature $T = 1.6$ K [1], yield enhanced coefficients $\gamma$ of the linear term in the electronic specific heat. The $\gamma$-values amount to 92 and 94 mJ/(mol$_U$)K$^2$ for $U_3Ni_3Sn_4$ and $U_3Pt_3Sn_4$ and to 380 and 280 mJ/(mol$_U$)K$^2$ for $U_3Cu_3Sn_4$ and $U_3Au_3Sn_4$, respectively (mol$_U$ denotes a mole uranium atoms). Heavy-electron behaviour in uranium intermetallics is often related to the proximity to a magnetic instability [2]. Indeed, $U_3Cu_3Sn_4$ is a heavy-electron antiferromagnet, with a relatively low Néel temperature $T_N = 12$ K [1]. Although magnetic order has not been reported for the other compounds, it is conceivable that expanding the lattice results in the emergence of magnetic order. For instance, by comparing the lattice parameters of $U_3Ni_3Sn_4$ ($a = 9.380$ Å) and $U_3Cu_3Sn_4$ ($a = 9.522$ Å), one may anticipate the existence of a magnetic quantum critical point (QCP) in the $U_3(Ni,Cu)_3Sn_4$ series. The possible existence of a QCP in this series is supported by the observation of an extended non-Fermi liquid (NFL) regime in the thermal, magnetic and transport properties of pure $U_3Ni_3Sn_4$ [3,4]. The absence of (weak) magnetic order in $U_3Ni_3Sn_4$, at least down to $T = 2$ K, was recently confirmed by µSR experiments [5].

Materials with low-temperature properties that show strong departures from the standard Fermi-liquid behavior nowadays attract considerable attention [6-8]. Several scenarios which may lead to NFL behavior in dense $f$-electron systems have been proposed. A first scenario is the proximity to a magnetic quantum critical point [9]. Here a transition takes place at $T = 0$ K from a magnetically ordered to a non-ordered state, either spontaneously (like in CeNi$_2$Ge$_2$ [10] and U$_2$Pt$_2$In [5,11]) or by tuning the transition temperature to 0 K by an external parameter, such as hydrostatic pressure (like in CePd$_2$Si$_2$ and CeIn$_3$ [12]) or chemical pressure (like in CeCu$_{6-x}$Au$_x$ [8]). In this case the thermodynamic properties are determined by collective modes corresponding to fluctuations of the order parameter in the vicinity of the critical point. A second possibility is a distribution of Kondo temperatures [13], where the Kondo effect on each $f$-electron impurity sets a different temperature scale, resulting in a broad range of effective Fermi temperatures (like in UCu$_{5-x}$Pd$_x$ [14]). Averaging over such a distribution gives rise to thermodynamic properties which follow NFL temperature dependencies. More recently, a third scenario has been proposed, namely the Griffiths phase model [15], where rare magnetic clusters favored by disorder appear in the vicinity of a quantum critical point. In the latter two models substantial disorder is a major ingredient. Since the single crystals used for this investigation have residual resistivity values $\rho_0$ of the order of 7 $\mu\Omega$cm, $U_3Ni_3Sn_4$ is a relatively clean material and the physics should not be dominated by disorder. This seems to rule out the Kondo-disorder and Griffiths phase models as possible causes for the NFL behavior reported for $U_3Ni_3Sn_4$.

For $U_3Ni_3Sn_4$, evidence for the proximity to an antiferromagnetic instability is predominantly provided by specific-heat experiments [3,4]. When analysing the low-temperature specific heat of $U_3Ni_3Sn_4$ three terms have to be considered: an electronic term $c_{el}$, a nuclear term $c_N$ due to Sn isotopes, which in the relevant temperature range is given by the high-temperature tail of a Schottky term ($c_N \sim 1/T^2$), and a lattice term $c_L \sim T^3$. Measurements of the specific heat on a single-crystalline sample in the temperature range $T = 0.3\text{-}5$ K [3] revealed the presence of a NFL electronic term of the form $c_{el}/T \sim \gamma_0 - \alpha T^{1/2}$. A $\alpha T^{1/2}$ correction to the standard Fermi-liquid coefficient has been predicted for a 3D antiferromagnetic zero-temperature critical point by Millis [9,16]. The same expression was derived by Moriya and Takimoto, within a self-consistent renormalization theory of spin fluctuations [17]. However, a drawback of the analysis presented in Ref.3, is the substantial nuclear Schottky term needed to make an adequate fit to the specific-heat data at the lowest temperatures. This resulted in a rather large calculated hyperfine field at the Sn site. Subsequent measurements of the specific heat down to lower temperatures (0.1 K) [4], showed that the electronic specific heat below $T = 0.4$ K is best described by the modified Fermi-liquid expression $c_{el}/T \sim \gamma_0 + \delta T^3 \ln(T/T^*)$, with $\gamma_0 = 0.130$ J/(mol$_U$)K$^2$. The $T^3\ln(T/T^*)$ term accounts for spin fluctuations with $T^* \sim 10$ K a characteristic spin-fluctuation temperature. In addition, a much smaller nuclear specific heat and an appropriate value for the hyperfine field ($B_{hf} = 0.5$ T) were found. Therefore, the analysis of the specific-heat data leads to the conclusion that $U_3Ni_3Sn_4$ has a Fermi-liquid ground state, with a crossover to non-Fermi liquid behavior near $T \sim 0.4$ K. The NFL behavior $c_{el}/T \sim \gamma_0 - \alpha T^{1/2}$, which pertains up to several Kelvin, suggests that $U_3Ni_3Sn_4$ is close to an antiferromagnetic quantum phase transition. NFL-like temperature dependencies have also been observed in the magnetic and transport properties [3]. The magnetic susceptibility varies as $\chi \propto T^{-0.3}$ for $T = 1.7\text{-}10$ K and the electrical resistivity varies as $\rho \propto T^{1.79}$ for $T$ = 1.7-12 K. However, as clearly illustrated by the specific-heat data [4], measurements of $\chi(T)$ and $\rho(T)$ down to much lower temperatures are needed in order to determine the true asymptotic behavior $(T \to 0)$.

Here we report measurements of the electrical resistivity of $U_3Ni_3Sn_4$ under hydrostatic pressures up to 1.8 GPa. In general, the application of pressure on non-magnetic heavy-fermion compounds drives the material further away from the magnetic instability. Within the simple Doniach approach [18] this can be understood from the increase of the Kondo energy, $k_BT_K$, with respect to the Ruderman-Kittel-Kasuya-Yosida interaction energy, $k_BT_{RKKY}$, due to the increase of the exchange parameter $J$ under pressure. Concurrently, the temperature $T_{FL}$ up to which the Fermi-liquid term in the resistivity, $\rho(T) \sim AT^2$, is observed should increase with pressure. As we will show below for $U_3Ni_3Sn_4$ $T_{FL} = a\,(p-p_c)^{1/2}$ where $p_c$ = -0.04±0.02 GPa is a critical pressure. Our data are consistent with $U_3Ni_3Sn_4$ exhibiting an antiferromagnetic QCP at $p_c$ = -0.04±0.02 GPa, as follows from an analysis within the theory of Rosch [19] for magnetotransport in metals with weak disorder close to an antiferromagnetic QCP.

The electrical resistivity of single-crystalline $U_3Ni_3Sn_4$ under pressure ($p \leq 1.8$ GPa) was measured on a bar-shaped sample in the temperature range 0.3-300 K. Since the crystal structure is cubic (space group I-43d) the resistivity is isotropic and the current could be applied along an arbitrary direction. Data were taken using a standard low-frequency four-probe ac-technique with a typical excitation current of ~ 100 μA. The value of the resistivity at room temperature $\rho_{RT}$ amounts to 380 μΩcm with an accuracy of 10% due to the uncertainty in the geometrical factor. This value of $\rho_{RT}$ is in very good agreement with the value of 385 μΩcm obtained on a polycrystal reported in Ref.1. The residual resistance ratio (RRR), $\rho_{RT}/\rho_0$, amounts to 55. For sample preparation and characterization we refer to Ref.3. Measurements under pressure were carried out using a copper-beryllium clamp cell. The sample was mounted on a specially designed plug and inserted into a teflon holder together with the pressure transmitting medium. A short tungsten carbide piston is used to transfer the pressure to the teflon holder. A mixture of Fluorinerts was used as pressure transmitting medium. The pressure values (accuracy 0.05 GPa) were calculated from the external load and corrected for an empirically determined efficiency of 80%. The pressure dependence of $\rho_{RT}$ was negligible.

In Fig.1 we show $\rho(T)$ of $U_3Ni_3Sn_4$ at ambient pressure and under pressures up to 1.8 GPa. The observed temperature variation of the resistivity at ambient pressure, i.e. a large value of $\rho_{RT}$, a weak maximum in $\rho(T)$ at a temperature $T_{max} \sim 240$ K and a strong decrease of $\rho(T)$ below $\sim 100$ K, is typical for dense Kondo systems. The temperature $T_{max}$ of the maximum in $d\rho/dT$, equals $15.8 \pm 0.5$ K, which provides a rough estimate of the coherence temperature, $T_{coh}$. Under pressure $T_{max}$ increases and equals $20.4 \pm 0.5$ K at 1.8 GPa. The qualitative behavior of $\rho(T)$ does not change in the range of pressures applied.

In Fig. 2 we show the low-temperature resistivity plotted versus $T^2$ ($< 10$ K$^2$). Initially, $\rho_0$ shows a weak reduction as a function of pressure, but attains a constant value of $\sim 6.2$ μΩcm at the highest pressures. The temperature range of the $\rho = \rho_0 + AT^2$ behavior becomes larger with pressure. In order to determine $T_{FL}$ at each pressure, we have fitted the data for increasing temperature ranges (keeping the lower bound fixed) to $\rho = \rho_0 + AT^2$, with $A$ as fit parameter and $\rho_0$ constant. By increasing the upper bound of the temperature interval of the fit, $A$ initially remains constant, but decreases when the upper bound exceeds $T_{FL}$. The values for $T_{FL}$, extracted by this fitting procedure, are indicated by the arrows in Fig.2, where the slopes of the dashed lines represent the resulting $A$-values. In Fig.3, we show the pressure variation of $T_{FL}$. Since the data extend down to 0.3 K only, the value $T_{FL} = 0.4 \pm 0.2$ K obtained at zero-pressure has a large error bar. However, the low-temperature specific-heat experiments yield the same value $T_{FL} = 0.4$ K [4].

Within the theory of Millis [9] for itinerant fermion systems $\rho(T) \propto T^{3/2}$ at the 3D antiferromagnetic quantum critical point. Away from the QCP the Fermi-liquid regime $\rho(T) \propto T^2$ grows. Using pressure as parameter which controls the distance to the QCP, the theory predicts $T_{FL} \sim (p-p_c)$, where $p_c$ is the critical pressure. However, very recently, a more detailed magnetotransport theory near a magnetic QCP has been presented by Rosch [19]. This theory delineates the NFL and Fermi-liquid regimes as a function of the distance to the QCP *and* the amount of disorder. Three different regimes are predicted for the resistivity $\Delta\rho = \rho - \rho_0$. For the three dimensional antiferromagnetic case ($d = 3$) these are: (I) $\Delta\rho \sim t^{3/2}$, (II) $\Delta\rho \sim tx^{1/2}$ and (III) $\Delta\rho \sim t^2 r^{-1/2}$. Here $t = T/\Gamma$ measures the temperature, with $\Gamma$ typically of the order of $T_{coh}$, $x = \rho_0/\rho_m \approx 1/RRR$ measures the amount of disorder, where $\rho_m$ is a typical high-temperature ($t \sim 1$) resistivity value, and $r \propto (\delta - \delta_c)/\delta_c$ measures the distance to the QCP in the paramagnetic phase, where $\delta_c$ is the critical control parameter. The temperature ranges of the different regimes depend on the amount of disorder in the system. For very clean samples ($x \ll 1$), the $\Delta\rho \sim t^{3/2}$ dependence is observed only at very low temperatures, and shows a cross-over to $\Delta\rho \sim tx^{1/2}$ there above. In the paramagnetic phase ($\delta > \delta_c$) the Fermi-liquid state (regime III) is the ground state, with $T_{FL}$ increasing as the distance to the QCP increases. The way this regime evolves depends strongly on the amount of disorder. For metals with very weak disorder the linear dependence $T_{FL} \sim (p - p_c)$ (using pressure as control parameter) is restricted to the immediate vicinity of the QCP, while at further distances to the QCP $T_{FL}$ shows a cross-over to the dependence $T_{FL} = a (p-p_c)^{1/2}$. The data obtained for our relatively clean sample of $U_3Ni_3Sn_4$ are consistent with the latter dependence. The solid line in Fig.3 shows the results of a fit of the data to the function $T_{FL} = a (p-p_c)^{\nu}$. The parameters extracted from this fit are $p_c = -0.04\pm0.02$ GPa, $\nu = 0.50\pm0.07$ and $a = 2.0\pm0.1$ KGPa$^{-\nu}$. Thus the analysis of the pressure variation of $T_{FL}$ within the magnetotransport theory of Rosch is consistent with $U_3Ni_3Sn_4$ being located close to an antiferromagnetic QCP, with the QCP located at a *negative* critical pressure of ~ -0.04 GPa. For $T > T_{FL}$ the theory predicts a cross-over to regime II, $\Delta\rho \sim tx^{1/2}$. This regime should be observable for $T_{FL} < T < \Gamma x^{1/2}$. With x= 0.018 and assuming $\Gamma \approx T_{coh} \approx 16$ K (see above) the relevant temperature range at ambient pressure is 0.4 K< $T$ < 2.1 K. However, the data do not allow to delineate the linear regime unambiguously, which is possibly due to large cross-over effects.

In Fig.4 we show the pressure dependence of the coefficient $A$ of the $T^2$ term. When the material is driven away from the magnetic instability $A$ decreases. The value of $A$ at zero pressure cannot be determined reliably from the experimental data. By simply extrapolating $A(p)$ to zero pressure we obtain the estimate $A(p=0) = 0.45\pm0.05$ $\mu\Omega$cm K$^{-2}$. With this value for the $A$ coefficient and $\gamma(p=0) = 0.13$ J/(mol$_U$)K$^2$ [4], we deduce a ratio $A/\gamma^2 \sim 27$ $\mu\Omega$cmK$^2$(mol$_U$)$^2$J$^{-2}$. This value differs considerably from the phenomenological value ~ 10 $\mu\Omega$cmK$^2$(mol$_U$)$^2$J$^{-2}$ for heavy-electron compounds reported by Kadowaki and Woods [20]. This is not surprising, because the simple Fermi-liquid relations $A \sim (T_{sf})^{-2} \sim (T_{coh})^{-2}$ (where $T_{sf}$ is the spin fluctuation temperature) and $\gamma \sim (T_{coh})^{-1}$ should no longer be valid in the immediate vicinity of a QCP. Moreover, the QCP controls the physics over a wide temperature range, which leads to NFL properties. From the reduction of $A$ under pressure, standard Fermi-liquid scaling predicts an increase of $T_{coh}$ of ~ 60 % at 1.8 GPa. However, the relative increase at 1.8 GPa extracted from the experimental data, using the assumption that the temperature $T_{max}$ of the maximum in $d\rho/dT$ yields a rough estimate of $T_{coh}$, is only 30 %.

In conclusion, we have measured the resistivity of the moderate heavy-fermion material $U_3Ni_3Sn_4$ under hydrostatic pressure up to 1.8 GPa. The experiments have been carried out on a single-crystalline sample with a residual resistivity of ~ 7 $\mu\Omega$cm, which indicates that $U_3Ni_3Sn_4$ is a relatively clean material. Under pressure the Fermi-liquid regime is rapidly recovered, as follows from the increase of the temperature $T_{FL}$, below which the resistivity obeys the Fermi-liquid expression $\rho \sim AT^2$. Concurrently, the coefficient $A$ decreases significantly. The measured pressure variation $T_{FL} \sim (p-p_c)^{1/2}$ lends strong support for an antiferromagnetic quantum critical point in $U_3Ni_3Sn_4$ at a negative critical pressure $p_c$ of -0.04±0.02 GPa, as deduced from the analysis within the magnetotransport theory of Rosch. This is consistent with the presence of the non-Fermi liquid term, $c_{el}/T \sim \gamma_0 - \alpha T^{1/2}$, previously reported in the low-temperature specific heat. Our results indicate that a relatively weak expansion of the lattice is sufficient to reach the QCP. We propose that pseudo-ternary $U_3(Ni,Cu)_3Sn_4$ compounds might form a suitable series to investigate the antiferromagnetic quantum phase transition.

**Acknowledgements**

P.E. acknowledges the European Commission for a Marie Curie Fellowship within the TMR program. The authors thank A. Matsushita for assistance in developing the high-pressure cell.

**Figure captions**

Fig.1 Resistivity as a function of temperature of single-crystalline $U_3Ni_3Sn_4$ at pressures of 0, 0.2, 0.6, 1.0, 1.4 and 1.8 GPa (curves from left to right).

Fig.2 Resistivity as a function of $T^2$ of single-crystalline $U_3Ni_3Sn_4$ at pressures of 0, 0.2, 0.6, 1.0, 1.4 and 1.8 GPa. The arrows indicate $T_{FL}$, i.e. the temperature up to which $\rho = \rho_0 + AT^2$. The slopes of the dashed lines yield the coefficients $A$. For $p = 0$ the dashed line is based on an extrapolation of the data under pressure.

Fig.3 $T_{FL}$, deduced from the resistivity, as a function of pressure for $U_3Ni_3Sn_4$. The solid curve represents a fit of the data (open circles) to the function $T_{FL} = a (p - p_c)^\nu$ with $\nu = 0.50 \pm 0.07$ and a critical pressure $p_c = -0.04 \pm 0.02$ GPa (see text). The dashed line indicates schematically the possible location of an antiferromagnetic (AF) phase boundary for $p < p_c$.

Fig.4 The coefficient $A$ as a function of pressure for $U_3Ni_3Sn_4$. The solid curve is to guide the eye.

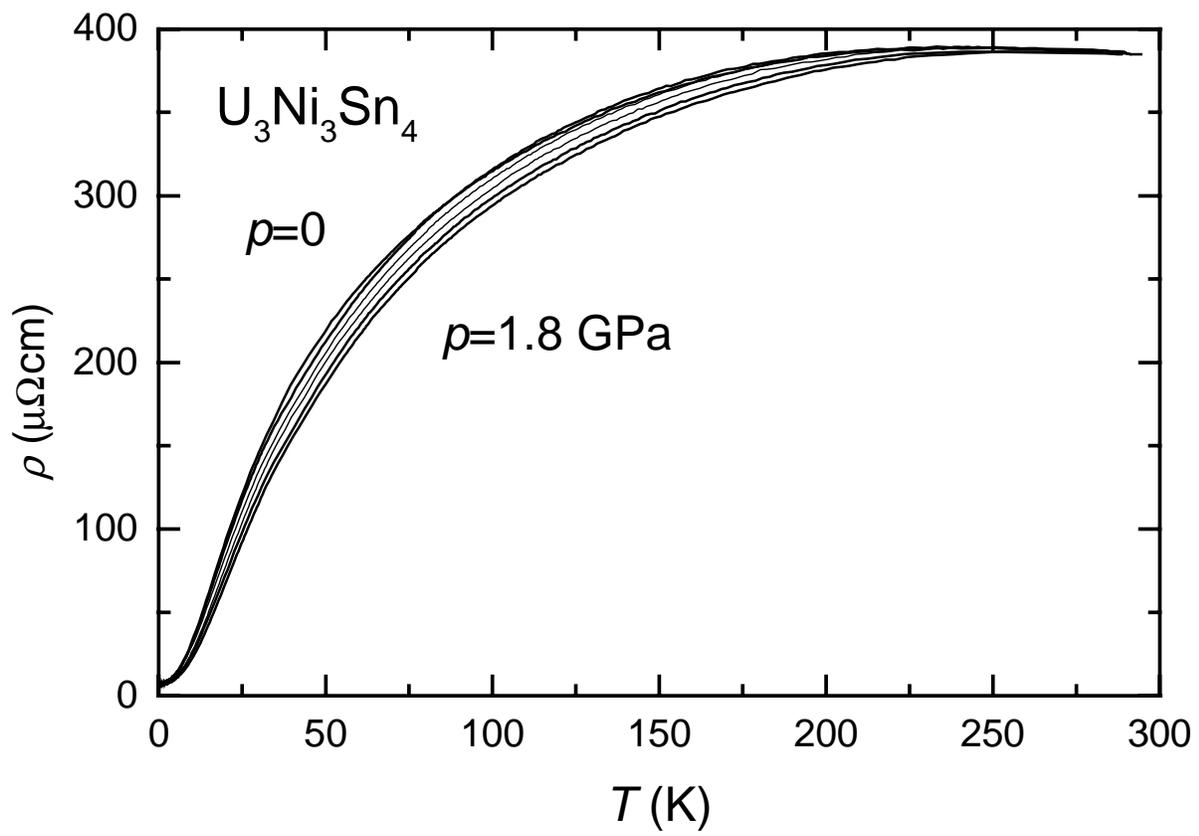

**Figure 1**

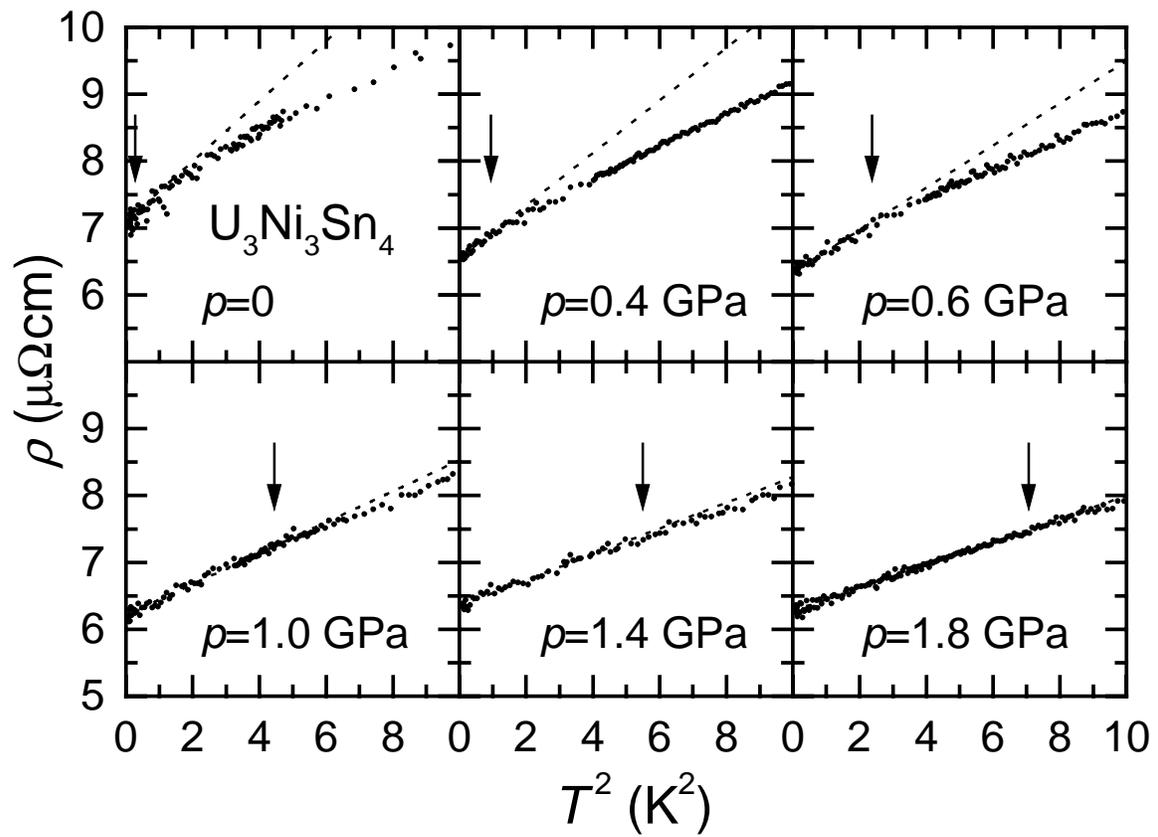

**Figure 2**

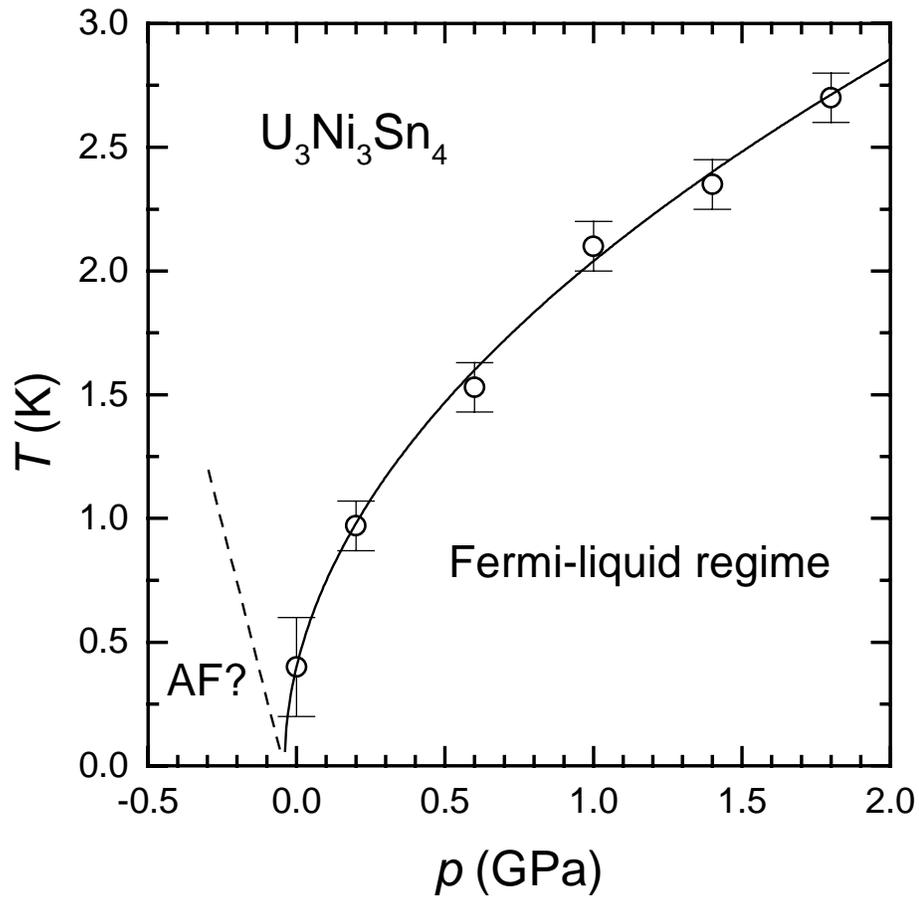

**Figure 3**

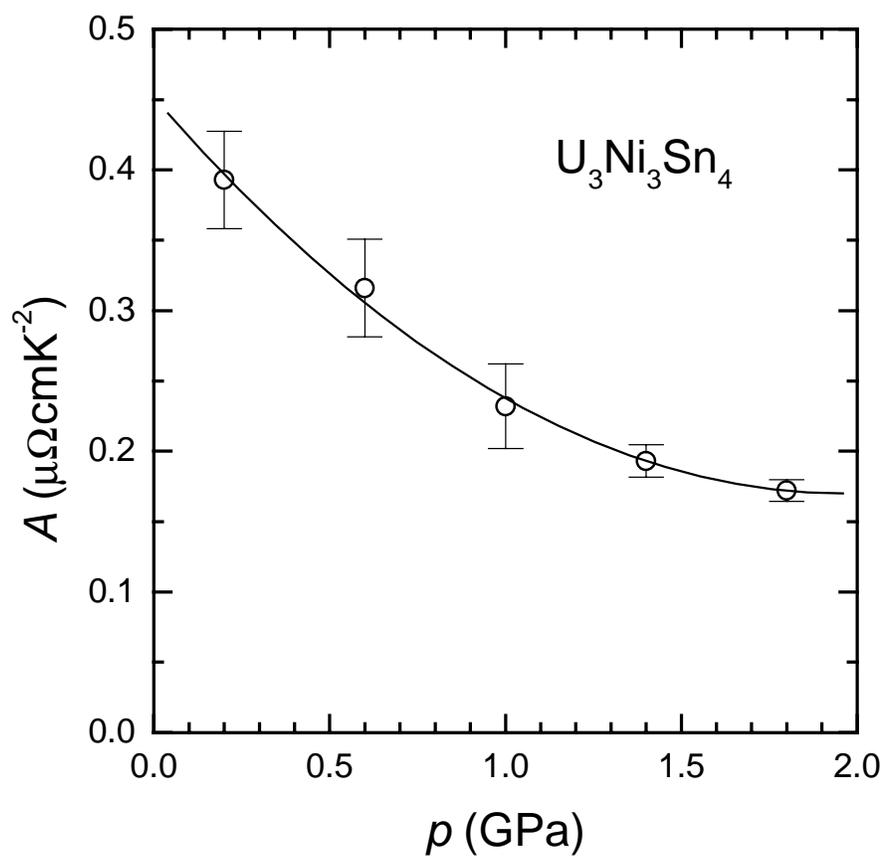

**Figure 4**